\documentclass[a4paper,11pt]{article}
\usepackage[utf8]{inputenc}
\usepackage{color}
\usepackage{xcolor}
\usepackage{amsmath}
\usepackage{amsthm}
\usepackage{adjustbox}
\usepackage{graphicx}
\usepackage{float}
\usepackage{multirow}
\usepackage[mathscr]{eucal}
\usepackage{subfigure}
\usepackage{epstopdf}
\usepackage{caption}
\usepackage{verbatim}
\usepackage{natbib}
\usepackage{colortbl}
 	\definecolor{manatee}{rgb}{0.59, 0.6, 0.67}

\usepackage{lineno,hyperref}

\usepackage{ulem}
\usepackage{setspace}
\usepackage[a4paper,
            left=1in,
            right=1in]{geometry}

\usepackage{authblk}

\title{Selecting the best compositions of a wheelchair basketball team: a data-driven approach}
\author[1]{Gabriel Calvo}
\author[1]{Carmen Armero}
\author[2]{Bernd Grimm}
\author[3]{Christophe Ley}

\affil[1]{Department of Statistics and Operations Research, Universitat de Val\`encia}
\affil[2]{Department of Precision Health, Luxembourg Institute of Health}
\affil[3]{Department of Mathematics, University of Luxembourg}
\date{September 2023}

\begin{document}

\maketitle

\begin{abstract}
Wheelchair basketball, regulated by the International Wheelchair Basketball Federation, is a sport designed for individuals with physical disabilities. This paper presents a data-driven tool that effectively determines optimal team line-ups based on past performance data and metrics for player effectiveness. Our proposed methodology involves combining a Bayesian longitudinal model with an integer linear problem to optimise the line-up of a wheelchair basketball team.  To illustrate our approach, we use real data from a team competing in the Rollstuhlbasketball Bundesliga, namely the Doneck Dolphins Trier. We consider three distinct performance metrics for each player and incorporate uncertainty from the posterior predictive distribution of the longitudinal model into the optimisation process. The results demonstrate the tool's ability to select the most suitable team compositions and calculate posterior probabilities of compatibility or incompatibility among players on the court.
\end{abstract}

\section{Introduction}

{Sports analytics is an important topic in applied statistics that has gained significance in recent years \citep{albert2001baseball,albert2017handbook,ley2020science}, particularly in basketball. The introduction of box scores by Henry Chadwick in the 1900s \citep{terner2021modeling} marked a milestone in the field. Notably, \citet{manley1989martin} created the first-ever player evaluation metric known as individual player efficiency. According to \citet{zuccolotto2020basketball}, \citet{oliver2004basketball} was a pioneer in the context of player and game analysis in this sport. Meanwhile, \citet{hollinger2005pro} developed the performance efficiency rating (PER), which is now commonly used in several sports media. The incorporation of tracking data in 2010 \citep{terner2021modeling} has significantly advanced basketball analytics, providing temporal and spatial information on players and matches. However, these advancements have not yet extended to sports for individuals with disabilities.}

Wheelchair basketball is widely recognised as one of the most popular team sports for people with a physical disability. It is regulated by the International Wheelchair Basketball Federation (IWBF). According to its rules, a wheelchair basketball match is played on a standard basketball court with two teams consisting of five players each. The sport is based on a Player Classification Points System (PCPS) that rates each player according to his or her playing ability. The valid player classifications in the IWBF are 1.0, 1.5, 2.0, 2.5, 3.0, 3.5, 4.0 and 4.5, the last value being the highest possible and indicating a high level of capability in playing basketball. At no time in a match shall a team have players participating whose total points exceed the 14-point limit \citep{council2021official}.

Choosing the optimal line-up for a basketball match can be challenging in any sport, but it may be particularly difficult for sports that have a PCPS for players. In fact,  the majority of scientific research on sports for individuals with disabilities focuses on player classifications in order ``to guarantee the fairness of results and ensure equal opportunities for athletes with different types and grades of disability'' \citep{gil2010wheelchair}.

This paper presents  a statistical procedure to select the best line-up for a wheelchair basketball team based on individual metrics of the performance of its different players in a series of previous basketball matches.   The proposed methodology involves   three consecutive stages. The first two are statistical in nature, a first inferential process followed by a predictive one, that assess   the temporal performance of each player by means of a  Bayesian longitudinal model. The third uses mathematical procedures of integer linear programming and the information provided by the predictive stage  to find ideal team line-ups. Our proposal accounts for game constraints from the PCPS and the inherent uncertainty that arises from the inferential and predictive  processes. It is worth noting that the presence of Bayesian modelling in this proposal allows for the incorporation and assessment of uncertainty in all its stages. The procedure we present here  can be a  useful tool for coaches and coaching staffs of teams, not only for selecting the best alignment of a team but also for evaluating  the performance of their players and opponents  and identifying  the strengths and weaknesses of different line-ups. 
 
We implement the approach proposed  in a study of the  wheelchair basketball  Doneck Dolphins Trier team, which participates in the first division of the Rollstuhlbasketball-Bundesliga (RBBL), the German league. The IWBF allows the use of some specific rules in national leagues that may differ slightly from those of international leagues and competitions.   In the case of the  RBBL,  the teams consist of mixed genders, each female player receives a $1.5$ on-court points bonus. It should be noted that there is a limit to the total score on the court, which is $17.5$. Additionally, similar bonuses may also be granted to younger or beginner players. For our analysis, we focus on data from the 2022-2023 Regular Season and  consider three  individual performance measurements, namely individual player efficiency, performance index rating, and win score. These are basketball metrics commonly used in National Basketball Association (NBA) and
Euroleague games   for evaluating player performance \citep{2020_sarlis} that combine many   positive and negative factors of each player in each match. 

The remainder of this article is organised as follows. Section \ref{sec:methodology} introduces the  methodology proposed in this study. In Section \ref{sec:application}, we delve into the study of the  wheelchair basketball team Doneck Dolphins Trier in the 2022-2023 Regular Season, display the performance of its players with regard to the metrics considered, apply the Bayesian model and the relevant optimisation problem, and discuss the main results and some practical implications about the composition of the best teams, highlighting its effectiveness. Lastly, Section \ref{sec:conclusion} presents some conclusions drawn from our research and outlines key avenues for future research.





\section{Methodology}\label{sec:methodology}

\subsection{Bayesian longitudinal modelling}

Consider the random vector $\boldsymbol{y}_i = (y_{i1}, \ldots, y_{iM_i})$, where $y_{ij}$ describes the performance of player $i$, $i = 1, \ldots, N$, at basketball match (or time) $j$, $j = 1, \ldots, M_i$. We define $\boldsymbol{y} = (\boldsymbol{y}_1, \ldots, \boldsymbol{y}_N)$ {as the longitudinal random vector comprising the performance records of all players}. The probabilistic behaviour of $\boldsymbol y$ depends on a set $\boldsymbol{\theta}$ of parameters and hyperparameters  and a vector $\boldsymbol{b}$ of random effects  so that the joint distribution of these elements can be expressed as
\begin{equation}
\begin{split}
    f(\boldsymbol{y}, \boldsymbol{\theta},\boldsymbol{b}) &=
  f(\boldsymbol{y} \mid \boldsymbol{\theta},\boldsymbol{b})
f(\boldsymbol{b}\mid \boldsymbol{\theta})\pi(\boldsymbol{\theta})\\
&= \bigg[\prod_{i=1}^{N}  f(\boldsymbol{y}_i \mid \boldsymbol{\theta},\boldsymbol{b}) \bigg]
f(\boldsymbol{b}\mid \boldsymbol{\theta})\pi(\boldsymbol{\theta}),
\end{split}\label{eqn:joint}
\end{equation}

\noindent where $f(\boldsymbol y \mid \boldsymbol \theta, \boldsymbol b)$ is the conditional distribution of $\boldsymbol y$ given  $\boldsymbol{\theta}$ and  $\boldsymbol{b}$ (usually referred to as the sampling model),  $f(\boldsymbol b \mid \boldsymbol \theta)$ is the conditional distribution for the random effects  $\boldsymbol b$ given  $\boldsymbol{\theta}$,  and  $\pi(\boldsymbol{\theta})$ is the prior distribution for  $\boldsymbol \theta$.
  Conditional independence of the response variable between  individuals is   assumed. For this reason, the conditional distribution $f(\boldsymbol y \mid \boldsymbol \theta, \boldsymbol b)$ can be expressed as the product of the conditional distributions associated with the different individuals in the study. Note that the prior distribution $\pi(\boldsymbol{\theta})$ is an indispensable element  within the  formulation of this Bayesian model.

The selection of the appropriate model for $f(\boldsymbol{y}_i \mid \boldsymbol{\theta},\boldsymbol{b})$ can vary from issue to issue depending on the nature of the problem.  The most common approach for analysing this type of repeated measurements data is via Mixed Linear Models, which assume normally  distributed scenarios  \citep{weiss2005modeling}. Alternatively, Generalized Linear Mixed Models can be employed when the behaviour of the response variables are within the framework of the exponential family of distributions, for example when the response variable is binary and logistic regression is employed for analysis \citep{diggle2002analysis}. 

Bayesian statistics combines the information prior to the data and the information provided by the data through Bayes' theorem. The probability distribution containing both information is the  posterior distribution 
 $\pi(\boldsymbol \theta, \boldsymbol b \mid  \mathcal{D})$, obtained through Bayes' theorem in terms of the likelihood function  $\mathcal L (\mathcal D \mid \boldsymbol \theta, \boldsymbol b)$ and the prior information as follows

 $$\pi(\boldsymbol \theta, \boldsymbol b \mid  \mathcal{D}) \propto \mathcal L( \mathcal D \mid \boldsymbol \theta, \boldsymbol b) \, f(\boldsymbol{b}\mid \boldsymbol{\theta}) \,\pi(\boldsymbol{\theta}).$$

{\noindent Here, the data $\mathcal{D}$ corresponds to observations of the random variable $\boldsymbol{y}$, and the symbol $\propto$ means ``proportional to" since the posterior distribution is equal to the product of these terms divided by an unknown constant, which is known as the marginal likelihood or evidence, denoted as $m(\mathcal{D})$.}
{Finally, the Bayesian model is completed by specifying a prior distribution, denoted as $\pi(\boldsymbol{\theta})$, for the model's parameters and hyperparameters. Our approach in this study is to adopt a minimally informative prior scenario.}

 The   posterior distribution is generally not analytical and needs to be approximated numerically. Markov Chain Monte Carlo (MCMC) methods  are the most popular numerical procedures for obtaining an approximate sample from the posterior distribution. {For that reason, in this work, we approximate the posterior distribution of the parameters by using MCMC methods.}

\subsection{Prediction} \label{subsec:pred}
Prediction is always associated with the results of a hypothetical new experiment. In our case, we are interested in predicting the response $\boldsymbol{y}^{(pre)} = (y_1^{(pre)}, \dots, y_N^{(pre)})$ of each player $i=1,\ldots,N$ in a hypothetical future sport event.  In the Bayesian framework, this objective is defined by the  posterior predictive distribution, computed from the sampling model and the posterior distribution $\pi(\boldsymbol \theta, \boldsymbol b \mid  \mathcal{D})$  as follows:

\begin{equation}\label{eqn:predictive}
  f(\boldsymbol{y}^{(pre)}  \mid  \mathcal{D}) =  
    \int \, f(\boldsymbol{y}^{(pre)} \mid \boldsymbol \theta, \boldsymbol b) \, \pi(\boldsymbol \theta, \boldsymbol b \mid  \mathcal{D})\, \mbox{d}(\boldsymbol \theta, \boldsymbol b).
\end{equation}

Since this predictive distribution depends on the  posterior distribution $\pi(\boldsymbol \theta, \boldsymbol b \mid  \mathcal{D})$, a usually non-analytic distribution of which we will have an approximate sample, we can use it by employing the Composition method \citep{tanner2012tools} to obtain   an approximate sample of the predictive distribution.  

The Composition method involves sampling $S$ simulations, $\{x^{(1)},\ldots,\\ x^{(S)}\}$, from an unknown probability density $h(x) = \int h_2(x|z) h_1(z) \, \mbox{d}z$, using the known densities $h_1(z)$ and $h_2(x|z)$. Initially, the method simulates $\{z^{(1)},\ldots,z^{(S)}\}$ from $h_1(z)$, and then computes $\{x^{(1)},\ldots,x^{(S)}\}$ from $h_2(x|z^{(s)})$ for $s=1,\ldots,S$.

{Thus, applying this idea, a sample of the posterior predictive distribution can be simulated by means of the following algorithm. For $s=1$ to $S$, repeat the following steps:

\begin{enumerate}
    \item Generate an approximate observation $\{\boldsymbol{\theta}^{(s)}, \boldsymbol{b}^{(s)}\}$ from $\pi(\boldsymbol \theta, \boldsymbol b \mid  \mathcal{D})$ through MCMC methods.
    \item Generate ${\boldsymbol{y}^{(pre)}}^{(s)}$ by simulating from $f(\boldsymbol{y}^{(pre)} \mid \boldsymbol{\theta}^{(s)}, \boldsymbol{b}^{(s)})$.    
\end{enumerate}
\noindent In this manner, we have computed an approximate observation ${\boldsymbol{y}^{(pre)}}^{(s)}$ from the posterior predictive distribution $({\boldsymbol{y}^{(pre)}}^{(s)} \mid \mathcal{D})$  shown in (\ref{eqn:predictive}).
 }

\subsection{Selecting  the optimal team line-up }

The ultimate goal of this paper is to propose a procedure for selecting the optimal team   based on the predicted behaviour of the different players of the team in a future match. This objective involves a stochastic  integer linear programming optimisation problem whose goal is to find the set of five possible players on the field that maximises their team's performance. This optimisation problem can be defined    by the maximisation of the objective function
\begin{equation}
    \text{max } \sum_{i = 1}^{N} z_i {y_i^{(pre)}},
    \label{eqn:objective}
\end{equation}
\noindent where $z_i$ indicates the presence of the player $i$ in the team combination, i. e., this variable will be $1$ if the player $i$ is included in the line-up team and $0$ otherwise. Again, $y_i^{(pre)}$ is the prediction of player $i$ in a future basketball match.

This problem contains two   constraints. First, just five players of the same team can play together on the court. The mathematical expression for this condition is

\begin{equation}
\label{eqn:restriction1}
   \sum_{i = 1}^{N} z_i = 5.
\end{equation}

\noindent The second constraint, which characterises wheelchair basketball, is related to the PCPS. As noted above, the maximum  punctuation value of the five players allowed on the field  by the IWBF is 14. Thus, this general restriction can be expressed as the inequality constraint

\begin{equation}\label{eqn:restriction2}
   \sum_{i = 1}^{N} C_i z_i \leq 14,
\end{equation} 

\noindent where $C_i$ is the ``class'' of player $i$, i.e., the functional punctuation of this player. 

The nature of the objective function in (\ref{eqn:objective}) is random because it depends on the posterior predictive distribution (\ref{eqn:predictive}). Consequently, the solution of the optimisation process will also be random in nature, in particular, we will get a posterior predictive distribution for all possible line-ups of the team. 
{As previously mentioned, we obtain the posterior distribution using MCMC methods, i.e., by sampling from an approximate posterior distribution. Remember that the predictive distribution can be sampled by following steps 1 and 2 outlined in Subsection \ref{subsec:pred}. The next stage involves the following step at each iteration $s$:}

\begin{enumerate}
   \item[3] Use the predictive observation $\boldsymbol{y}^{(pre)(s)}$ to  solve the optimisation problem,  $\text{max } \sum_{i = 1}^{N} z_i \,{y_i^{(pre)(s)}}$ with constraints (\ref{eqn:restriction1}) and  (\ref{eqn:restriction2}), to obtain the subsequent line-up solution $\boldsymbol z^{(s)}=(z_1^{(s)}, \ldots,z_N^{(s)})$, where $\boldsymbol z^{(s)}$ will be a vector   of zeros and ones, with exactly five ones whose index will correspond to the players selected in the stage $s$ of the optimisation process.  
\end{enumerate}

\noindent The algorithm will provide a sample of size $S$ of the predictive distribution of the different ``best'' line-ups of the team. 

{In summary, the proposed methodology comprises three stages, as outlined in Steps 1, 2, and 3. Firstly, we derive the posterior distribution of the parameters for the longitudinal model. Next, we compute predictions for a new match using the posterior predictive distribution. Finally, we obtain the posterior distribution of solutions for the integer linear optimisation problem.
}

\section{Doneck Dolphins Trier case}\label{sec:application}
\subsection{Data and metrics for player effectiveness
}

  The methodology introduced above was applied to a wheelchair basketball team, the Doneck Dolphins Trier, which has been competing in the first division of the RBBL in Germany since 2003. The aim of this study was to determine optimal line-ups for the team based on the results of $18$ basketball matches of the season 2022-2023, obtained from the RBBL website (\url{http://stats.rollstuhlbasketball.de/}) and using only data from $9$  players who played more than 40 minutes during the season.

\begin{table}[ht]
		\centering
		\caption{Name, functional classification value and sex of the players of the Doneck Dolphins Trier team who play more than 40 minutes during the 2022-2023 RBBL season.}\label{tab:data}
		\vspace{0.25cm}
		\begin{tabular}{cccc}
			\noalign{\hrule height 1pt}
			\multicolumn{1}{c}{Player} & \multicolumn{1}{c}{Classification}    & \multicolumn{1}{c}{Sex} & Index \\ \noalign{\hrule height 1pt}
			Annabel Breuer   & \multicolumn{1}{r}{$1.5$} & \multicolumn{1}{r}{Woman} & \multicolumn{1}{r}{$1$}  \\
   			Correy Rossi   & \multicolumn{1}{r}{$2$} & \multicolumn{1}{r}{Man} & \multicolumn{1}{r}{$2$}  \\
      	Dejon Green   & \multicolumn{1}{r}{$3.5$} & \multicolumn{1}{r}{Man} & \multicolumn{1}{r}{$3$}  \\
            Dirk Passivan   & \multicolumn{1}{r}{$4.5$} & \multicolumn{1}{r}{Man} & \multicolumn{1}{r}{$4$}  \\
			Lucas Jung     & \multicolumn{1}{r}{$1$} & \multicolumn{1}{r}{Man} & \multicolumn{1}{r}{$5$}\\
			Natalie Passivan    & \multicolumn{1}{r}{$4.5$} & \multicolumn{1}{r}{Woman} & \multicolumn{1}{r}{$6$}\\
			Patrick Dorner     & \multicolumn{1}{r}{$3.5$} & \multicolumn{1}{r}{Man} & \multicolumn{1}{r}{$7$}\\
			Svenja Erni    & \multicolumn{1}{r}{$3.5$} & \multicolumn{1}{r}{Woman} & \multicolumn{1}{r}{$8$}\\
            Walter Vlaanderen   & \multicolumn{1}{r}{$4.5$} & \multicolumn{1}{r}{Man} & \multicolumn{1}{r}{$9$}\\\noalign{\hrule height 1pt}
		\end{tabular}
	\end{table}
 
Table \ref{tab:data} shows some general information for the nine players who met that criterion. Three of these players, Annabel Breuer, Natalie Passivan and Svenja Erni,  are women. Player Lucas Jung has a functional classification of $1$,  Annabel Breuer has $1.5$, and   Correy Rossi   has $2$ points.  Three players, Dejon Green, Patrick Dorner and Svenja Erni, have a functional classification of $3.5$, and Dirk Passivan, Natalie Passivan, and Walter Vlaanderen, have $4.5$ points, which is the maximum possible value indicating   a strong capacity to play basketball.

To carry out the analysis,  we have used three different measures of player performance. These are metrics commonly used in official competitions and sports studies to evaluate player performance.

\begin{itemize}
\item  \textbf{Individual player efficiency} (EFF). It is a full   effectiveness measure that attempts to assess a player's output over and above the number of points produced. It was developed by   Kansas City sports reporter and statistician Martin Manley \citep{manley1989martin} and it is one of the metrics commonly used by the  NBA.  EFF is computed by adding several positive events and subtracting some negative events as follows:
\begin{equation*}
\begin{split}
    \text{EFF} = &\text{(points + rebounds + assists + steals + blocks)} \, \,- \\
    &\text{(missed field goals + missed free throws + turnovers)}.
\end{split}
\end{equation*}

\item \textbf{Performance Index Rating} (PIR). This metric is similar to EFF but considers more actions   such as fouls drawn, shots rejected, and personal fouls. It is   employed in the Eurocup and many European leagues in the International Basketball Federation (FIBA) \citep{torres2016basketball}. PIR is defined as:

\begin{equation*}
\begin{split}
    \text{PIR} = &\text{(points + rebounds + assists + steals + blocks + fouls drawn)} \,\, - \\
    &\text{(missed field goals + missed free throws + turnovers +} \\
    &\ \text{shots rejected + personal fouls)}.
\end{split}
\end{equation*}

\item \textbf{Win Score}.  This metric was proposed by \citet{berri2010stumbling}. It substitutes missed field goals and missed free throws in EFF and PIR with field goals attempted and free throws attempted, respectively. In addition, Win Score weights some positive actions differently than EFF and PIR. It is defined as follows:
\begin{equation*}
\begin{split}
    \text{Win Score} = &\text{(points + rebounds + } 0.5 \,\text{assists + steals + } 0.5 \,\text{blocks)} - \\
    &\text{(field goals attempted + } 0.5 \,\text{free throws attempted + }\\ 
    &\text{turnovers + } 0.5 \,\text{personal fouls)}.
\end{split}
\end{equation*}

    \end{itemize}

\begin{figure}[p]
		\centering
		\subfigure[EFF per minute. ]{\includegraphics[width=105mm]{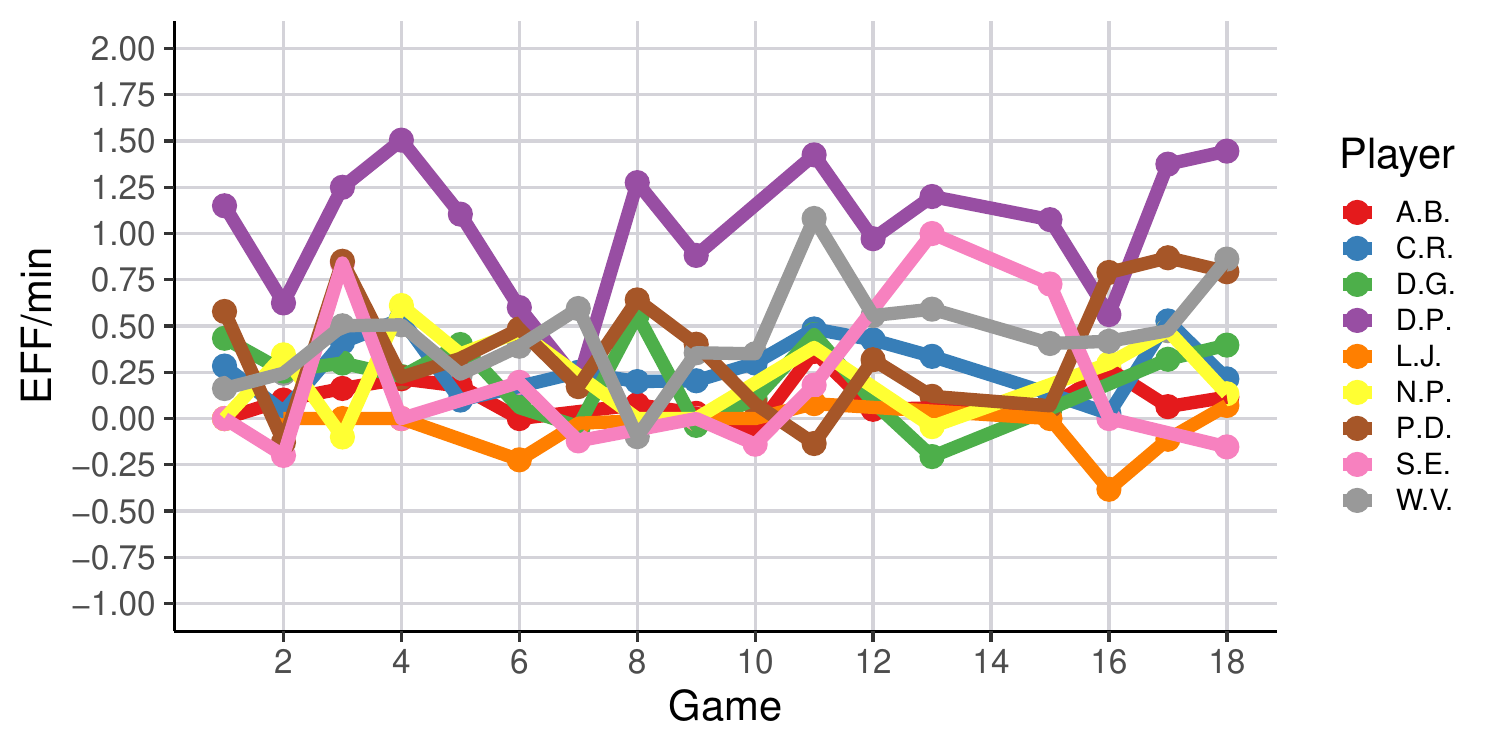}}
		\subfigure[PIR per minute.]{\includegraphics[width=105mm]{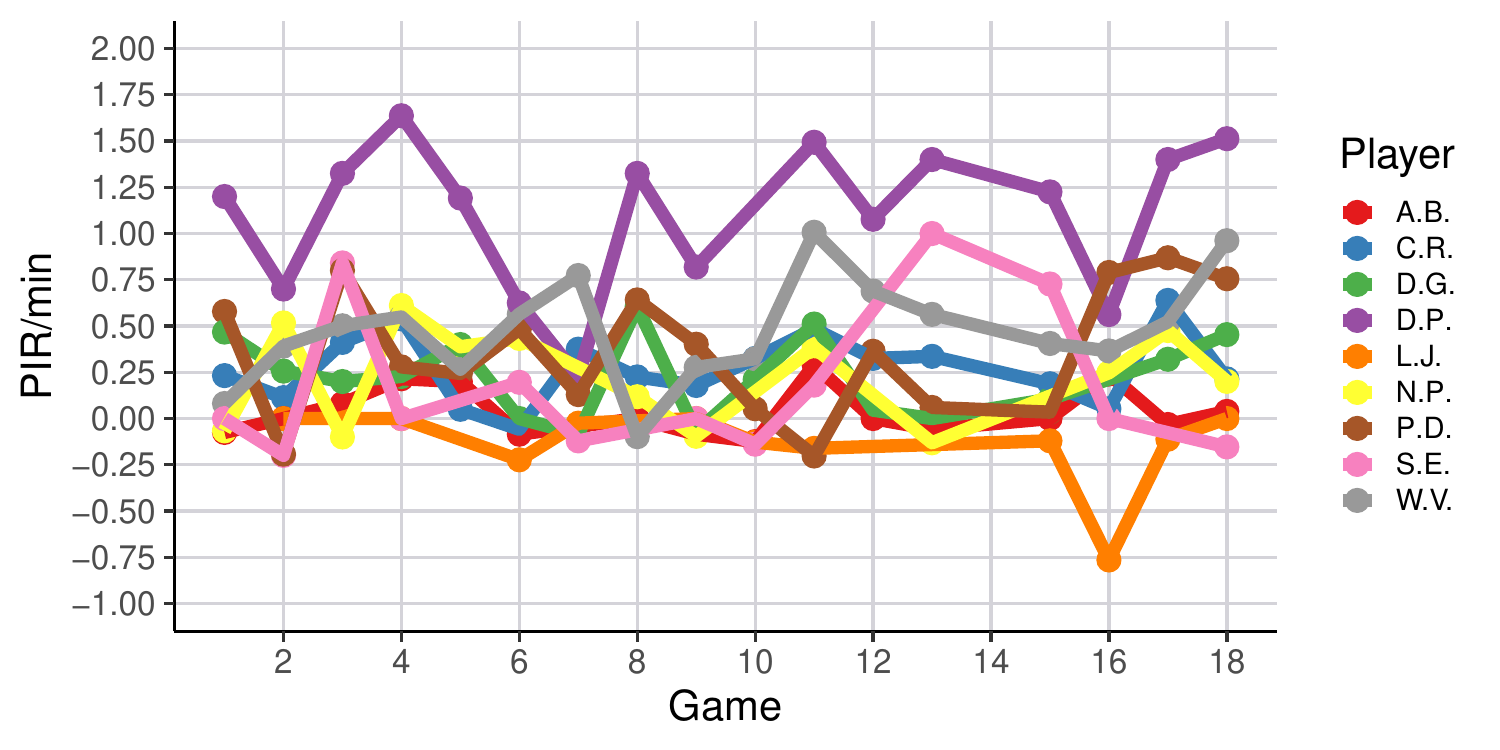}}	
        \subfigure[Win Score per minute.]{\includegraphics[width=105mm]{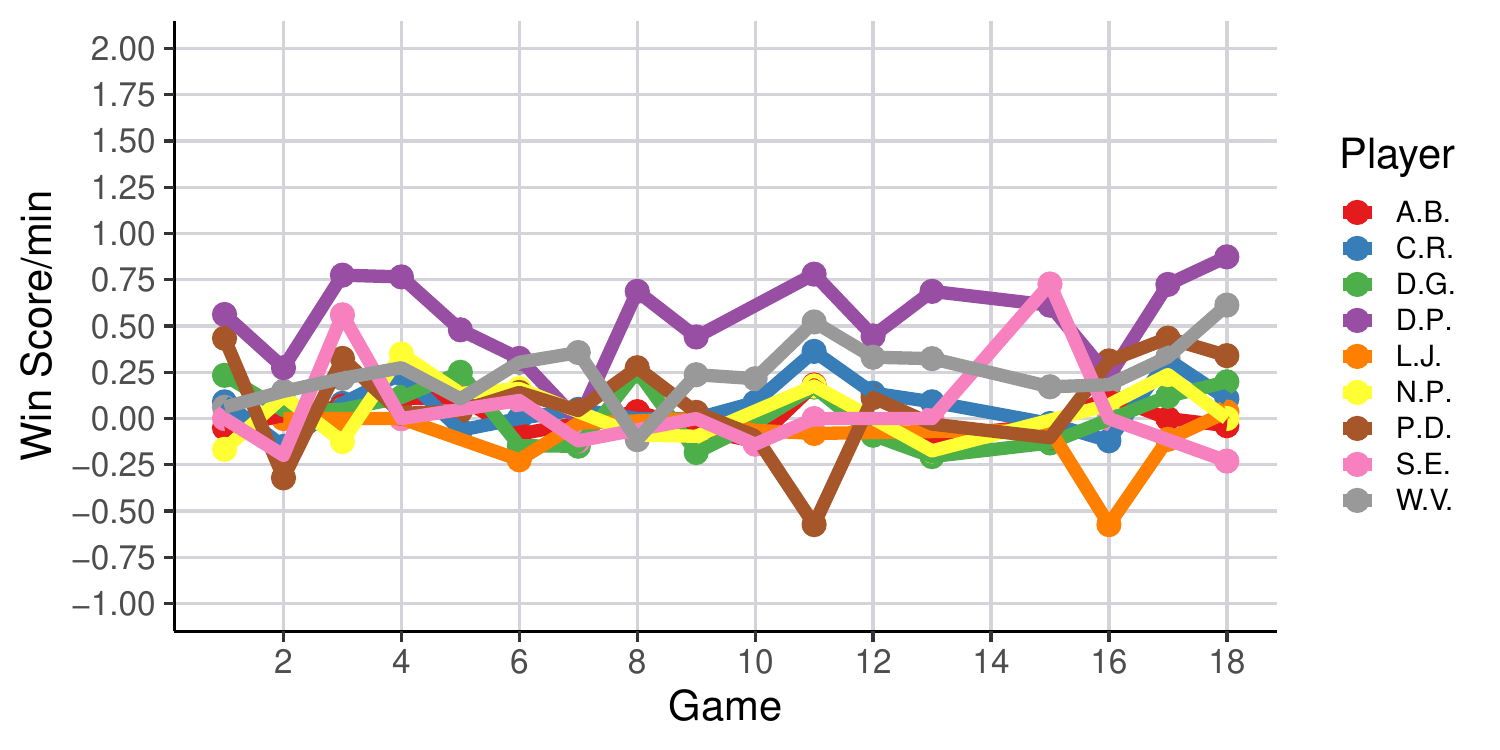}}
		\caption{Spaghetti plots of EFF, PIR and Win Score ratings per minute by player for the Doneck Dolphins Trier  team during the 2022-2023 RBBL Regular Season. Players are identified by the initials of their first and last name.}
		\label{fig:spaghetti}
	\end{figure}

EFF, PIR and Win Score have been computed for each player in each basketball match and divided by the number of minutes they played to assess their individual performance in every match.  EFF, PIR and Win Score observations per minute by player in each contest of the 2022-2023 season are presented in Figure \ref{fig:spaghetti}. The graphs of the three metrics show a behaviour with no general upward or downward patterns over the course of the matches. The variability of the EFF and mainly PIR observations of players in the same basketball match is quite similar and generally higher than that observed in Win Scores.  In all the metrics, the figure of Dirk Passiwan stands out. He has a rather irregular behaviour  but he performed exceptionally well compared to the other players, consistently ranking above them in most matches.  Walter Vlaanderen   also has good results. {Due to the restrictions imposed by the PCPS, it should be noted that a high match performance by a player does not necessarily imply their inclusion in the best line-up. In fact, it may be the case that an athlete with a low functional classification and relatively modest performance becomes one of the most frequently playing players. This decision by the coach can often be a sound one in many instances.
}

\subsection{Modelling of athlete performance}

We consider the longitudinal variable $y_{ij}$  representing the score of the player $i$, $i=1,\ldots,N=9$ at match $j$, $j=1,\ldots,M=18$.  We will assume   a sampling mixed linear model for $y_{ij}$ defined as follows
\begin{align}
(y_{ij}   \mid \boldsymbol{\theta}, \boldsymbol{b}) & \sim \mbox{N}(\mu_{ij}, \sigma^2) \nonumber\\
 (\mu_{ij} \mid  \boldsymbol{\theta}, \boldsymbol{b})  & = \beta_0 + b_{0i} + b_{0j} + 
    \beta_W I_W(i) + \beta_C C_i + \nonumber\\  & \;\;\;\; \beta_H I_H(j) + 
     (\beta_1 + b_{1i}) j ,
    \end{align}

\noindent where the parameter vector $\boldsymbol \theta$ contains all the parameters and hyperparameters of the model. {Parameters include the standard deviation parameter associated with the measurement error $\sigma$, and the common coefficients $\beta_0$, $\beta_W$, $\beta_C$, and $\beta_H$, which represent the common intercept and the coefficients associated with the covariates, respectively. Hyperparameters include the standard deviation parameters of the random effects $\sigma_0$, $\sigma_{0m}$, and $\sigma_1$.}
{Additionally,} $\boldsymbol{b}$ cointains all the random effects associated with players and matches. Individual random effects $b_{0i}$ and
random slopes $b_{1i}$ ($i=1,\dots,9$) are mutually independent and conditionally normally distributed as $(b_{0i}| \sigma_0^2)\sim \text{N} (0, \sigma_0^2)$ and $(b_{1i}| \sigma_1^2)\sim \text{N} (0, \sigma_1^2)$. Moreover, the random effect associated with the basketball match $b_{0j}$ ($j=1,\dots,18$)  is also normally distributed, $(b_{0j}| \sigma_{0m}^2)\sim \text{N} (0, \sigma_{0m}^2)$. We consider as covariates the gender indicator variable $I_W(i)$, $1$ when the player is female or $0$ otherwise, the functional classification, $C_i$, and a dummy variable $I_H(j)$ indicating if the match was played at home ($1$) or not ($0$).

To complete the Bayesian model we need to choose  a prior  distribution for all parameters and hyperparameters of the model. We assume a scenario of non-informative  and independent prior, and select a wide normal distribution for $\beta_0$, $\beta_1$ and for the coefficient parameters associated with the covariates, resulting in $\pi(\beta_0)=\pi(\beta_1)=\pi(\beta_W)=\pi(\beta_C)=\pi(\beta_H)= \text{N}(0, 10^2)$. Finally, uniform distributions for the standard deviation parameters $\pi (\sigma)=\pi (\sigma_0)=\pi (\sigma_1)=\pi (\sigma_{0m})=\text{U}(0, 10)$ are elicited.

We consider that the proposed Bayesian modelling is valid for the study of each of the three effectiveness metrics presented, EFF, PIR, and Win Score. Therefore, we will work with an inferential, predictive and optimisation process for each of them that will lead to different proposals for team alignments.

We approximate the subsequent  posterior distribution of $\boldsymbol{\theta}$ and $\boldsymbol{b}$ using  MCMC sampling methods \citep{tanner2012tools} through the JAGS software \citep{plummer2003jags}. Three parallel chains were executed for $100\hspace*{.35mm}000$ iterations each, preceded by burn-ins of $300\hspace*{.35mm}000$ iterations. In addition, to mitigate autocorrelation, the chains were thinned at every 100th iteration based on estimated autocorrelation in the sample to finally obtain  an approximate sample of size $S=3000$ from the target posterior distribution. 

In preparation for a potential upcoming basketball match after the completion of 18 matches in the season, our objective   is to find the optimal  team line-up  for this  future match according to each of the efficiency measures considered. From the approximate sample of the posterior distribution $\pi (\boldsymbol{\theta}, \boldsymbol{b} \mid \mathcal D)$
 for each efficiency measure, we have obtained, through (\ref{eqn:predictive}), a sample of the predictive distribution of the behaviour  $\boldsymbol{y}^{(pre)}$ of each of the team's players in the upcoming match. Each of these values has generated an integer programming problem which we have solved with the \textit{lpSolve} package \citep{lpSolve}. The complete analysis, performed using R code (version 4.0.5) \citep{Rsoft}, and the data are available as supplementary material at \url{https://github.com/gcalvobayarri/Wheelchair_basketball_lineups.git}.

 At this point, we would like to make a small point about the second constraint of the study's optimisation problem. 
 As already mentioned, the German RBBL has some particularities with respect to the international rules. To account for the RBBL  rules, we made the following  modifications to the class restriction constraint (\ref{eqn:restriction2}) in our optimisation problem with regard to the number of female players in the line-up team.

 \begin{enumerate}
 \item If there is no female player: \begin{equation} \label{eqn:cons1}\sum_{i = 1}^{N} C_i z_i \leq 14.5.
 \end{equation}

\item When there is just one female player: 
\begin{equation} \label{eqn:cons2}
   \sum_{i = 1}^{N} C_i z_i \leq 16.
\end{equation} 

\item In the case of two or more female players:
\begin{equation}\label{eqn:cons3}
   \sum_{i = 1}^{N} C_i z_i \leq 17.5.
\end{equation}
\end{enumerate}

\subsection{Selection of the best team}
The number of possible line-ups of the team with five players is 126, although not all of them will be valid because they will not meet constraints   (\ref{eqn:cons1}),  (\ref{eqn:cons2}), and  (\ref{eqn:cons3}).  In particular, out of the 6 possible line-ups with an all-male team only 2 are valid line-ups;  of the 45 potential line-ups with a team of four men and one woman, only 27  are acceptable; of the 60 possible teams with three men and two women only 48 meet the criterion (\ref{eqn:cons3}), and all of the 15 teams with three women and two men meet the above constraints. 
Therefore, the number of possible line-ups  is $92$ and we can estimate the posterior probability   that alignment $L_k,\ k=1,\dots,92$,  is the optimal team through the quotient between the number of optimisation times  that have resulted in   line-up $L_k$ and the number of optimisation problems solved.

According to the obtained results, the seven line-up teams with the highest probabilities are
\begin{equation*}
\begin{split}
    L_1 &= \{ \text{A. Breuer}, \text{C. Rossi}, \text{D. Passivan}, \text{P. Dorner}, \text{W. Vlaanderen} \},\\
    L_2 &= \{ \text{A. Breuer}, \text{D. Passivan}, \text{P. Dorner}, \text{S. Erni}, \text{W. Vlaanderen} \},\\
    L_3 &= \{ \text{A. Breuer}, \text{C. Rossi}, \text{D. Passivan}, \text{N. Passivan}, \text{W. Vlaanderen} \},\\
    L_4 &= \{ \text{A. Breuer}, \text{C. Rossi}, \text{D. Green}, \text{D. Passivan},  \text{W. Vlaanderen} \},\\
    L_5 &= \{ \text{A. Breuer}, \text{D. Green}, \text{D. Passivan},  \text{S. Erni}, \text{W. Vlaanderen} \},\\
    L_6 &= \{ \text{C. Rossi}, \text{D. Passivan},  \text{L. Jung}, \text{S. Erni}, \text{W. Vlaanderen} \},\\
    L_7 &= \{ \text{A. Breuer}, \text{C. Rossi}, \text{D. Passivan},  \text{S. Erni}, \text{W. Vlaanderen} \}.
\end{split}
\end{equation*}

\begin{figure}[p]
		\centering
		\subfigure[EFF per minute. ]{\includegraphics[width=90mm]{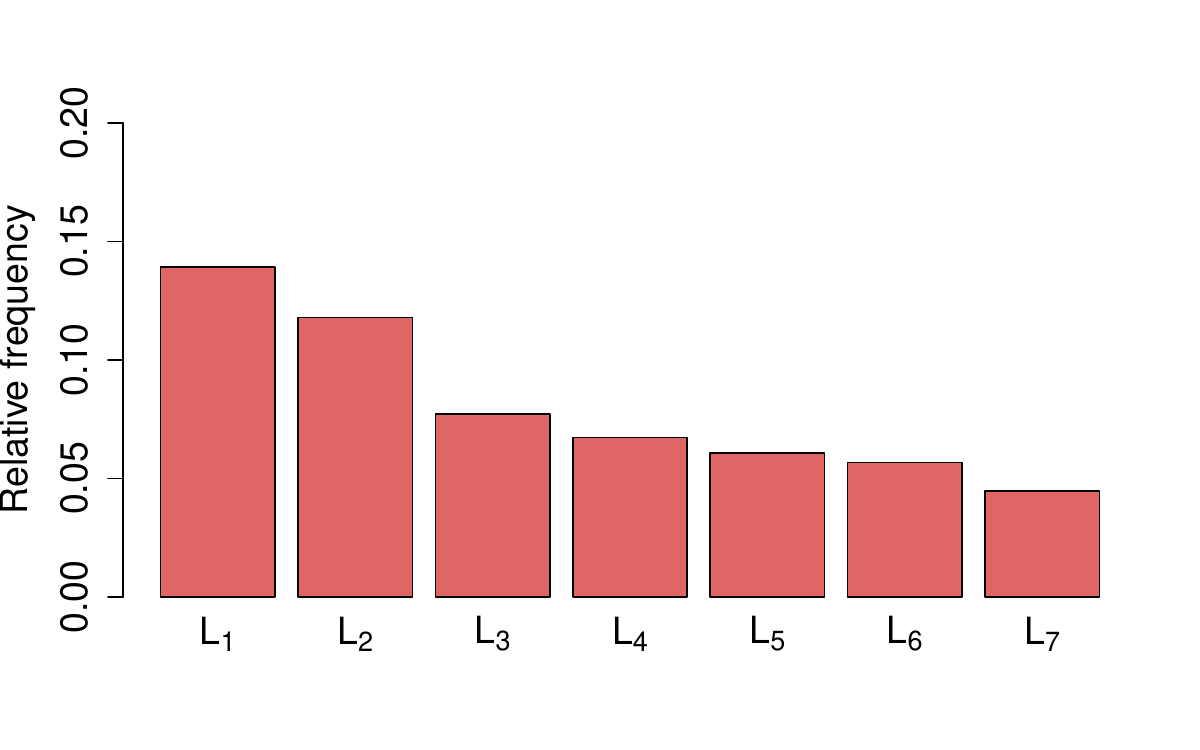}}
		\subfigure[PIR per minute.]{\includegraphics[width=90mm]{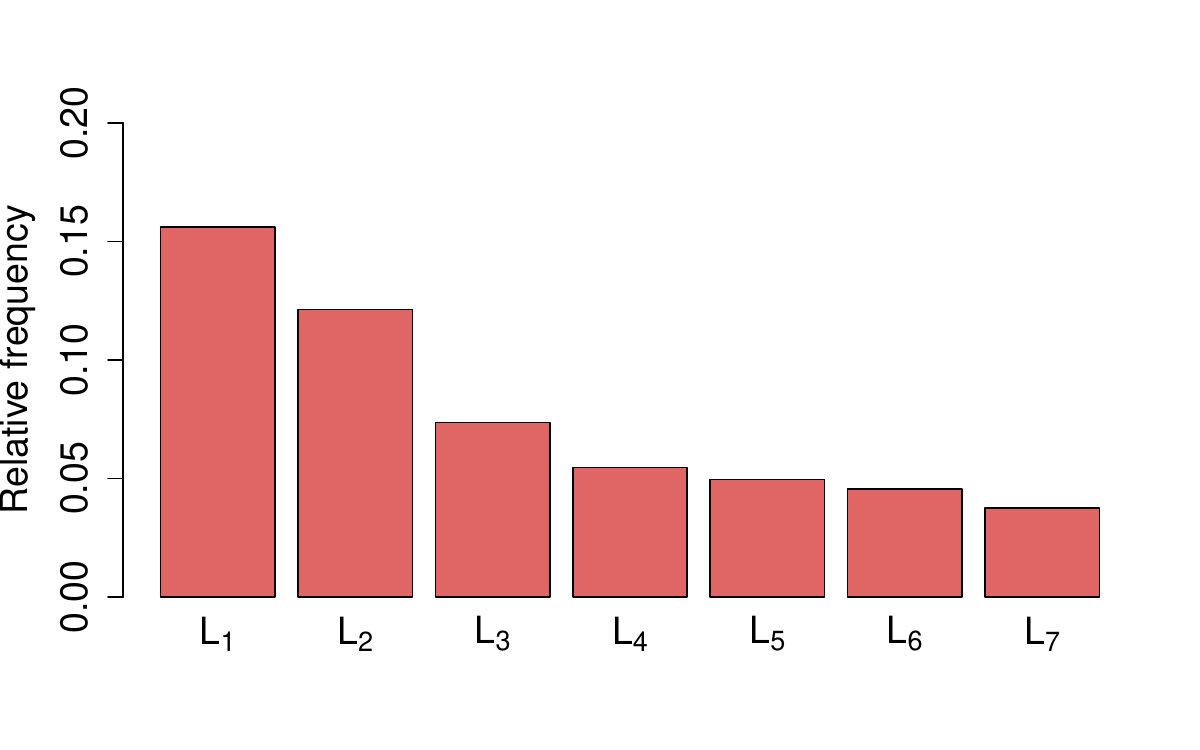}}	
        \subfigure[Win Score per minute.]{\includegraphics[width=90mm]{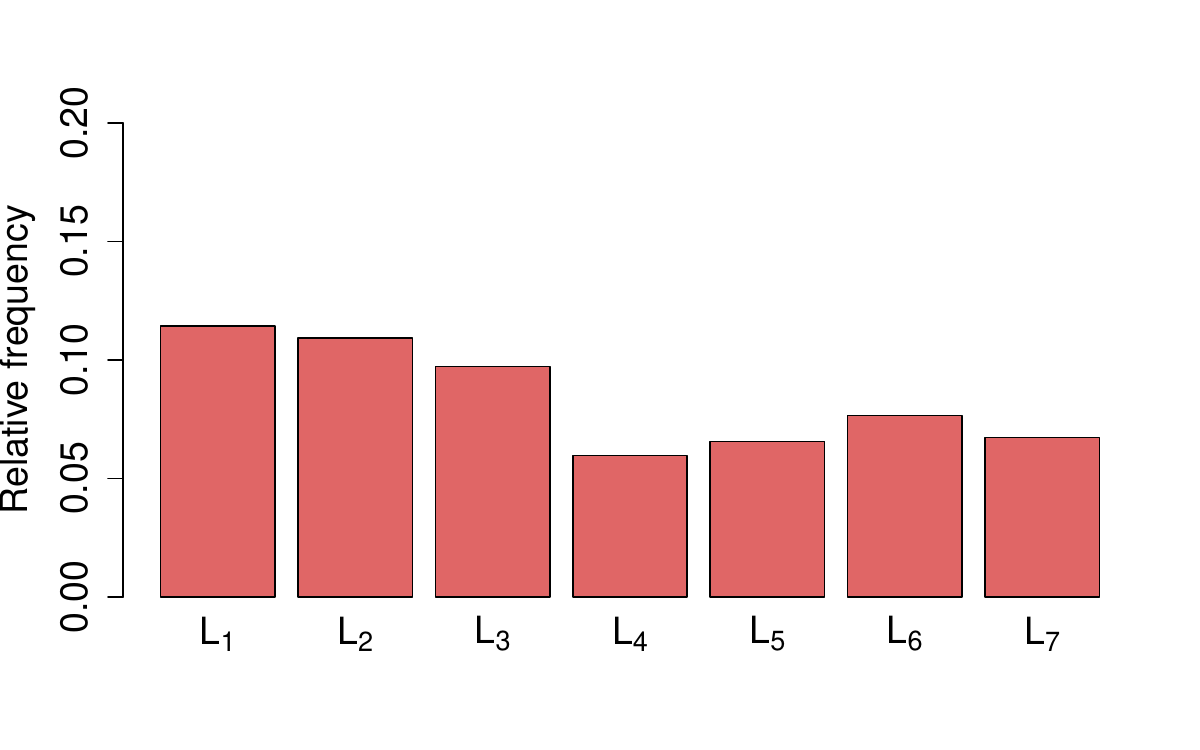}}
		\caption{Bar plots showing the estimated probabilities associated with being the most optimal line-ups for EFF, PIR and Win Score metrics for  compositions $L_1$, $L_2$, $L_3$, $L_4$, $L_5$, $L_6$, and $L_7$.   }
		\label{fig:barplot}
	\end{figure}

\noindent Bar plots in Figure \ref{fig:barplot} illustrate the posterior probabilities associated with these seven compositions which are identified as the most optimal line-ups for EFF, PIR and Win Score metrics. Each bar represents the posterior probability  that the corresponding alignment is the optimal. The remarkable thing that one can observe is that, based on the three performance metrics, $L_1$, $L_2$, and $L_3$ emerge as the top three optimal line-up teams, in that order. However, the ranking of $L_4$, $L_5$, $L_6$, and $L_7$ differs when considering the Win Score metric in comparison to EFF and PIR. Finally, it is worth noting that Dirk Passivan and Walter Vlaanderen appear in all the seven line-up teams, followed by Annabel Breuer who is included in six of the seven compositions.

\begin{figure}[H]
	\centering
	  \includegraphics[width=120mm]{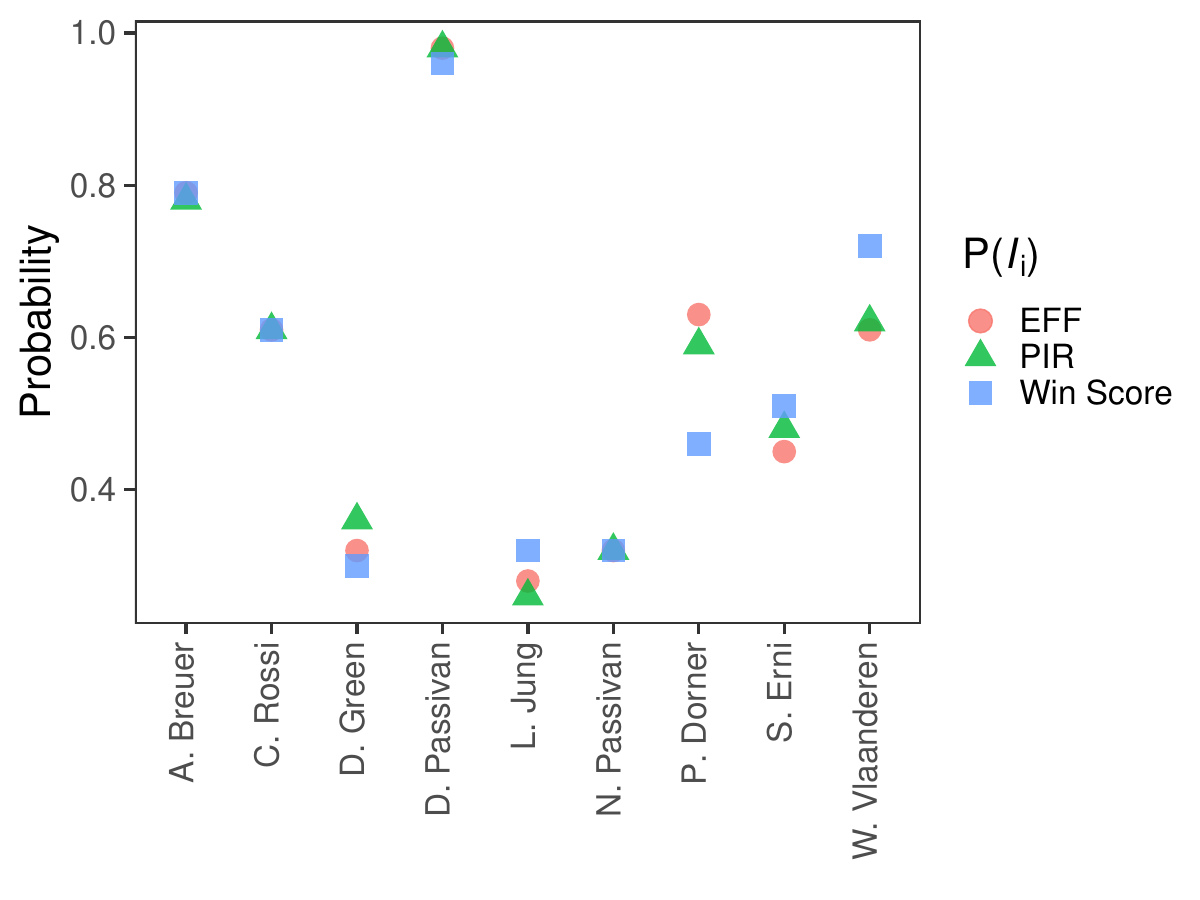} 
	 	\caption{Estimated probabilities $P(I_i \mid  \mathcal D)$ of player $i$ being included in the optimal line-up team based on EFF, PIR and Win Score metrics.} \label{fig:inclusion_probabilities}
\end{figure}

The results obtained allow us to add more insights to the exploration of the composition of the team's best line-ups. Clearly, in the same basketball match, it is extremely rare for the same team to play the whole time, as there are various changes during a game. These changes can occur due to different playing strategies depending on the evolution of the match, injured or suspended players, and other factors. In this regard, a valuable piece of information would be the probability of a player being included in the optimal team. We denote the event of including a specific player in the line-up as $I_{i}$, where $i=1,\dots,9$ corresponds to the indices of the players in Table \ref{tab:data}. Posterior probability   $P(I_i \mid \mathcal D)$ is estimated according to the number of times the player in question appears in the optimal team in relation to the number of optimisation problems addressed. Figure \ref{fig:inclusion_probabilities} shows these  probabilities  for each of   the three performance metrics   evaluated. Upon observing this graph, it becomes apparent that three players, namely Dejon Green, Lucas Jung, and Natalie Passivan, have the lowest probabilities of being included in the optimal composition. Conversely, four players stand out above the rest of the team: Annabel Breuer, Correy Rossi, Dirk Passivan, and Walter Vlaanderen.


We can also estimate the posterior probability of several players playing together. For example, the posterior  probability that both Annabel Breuer and Natalie Passivan are in the line-up that maximises the Win Score metric is given by:  
\begin{equation*}
    P(I_1, I_6 \mid \mathcal D) = 0.24.
\end{equation*}

\noindent Additionally, if we want to estimate the probability that these two players are included in the team composition, knowing that Correy Rossi is in the line-up, it is calculated as follows:
\begin{equation*}
    P(I_1, I_6 \mid I_2, \mathcal D) = \frac{P(I_1, I_6, I_2 \mid \mathcal D)}{P(I_2 \mid \mathcal D)}=\frac{0.23}{0.61}=0.38.
\end{equation*}

\begin{figure}[H]
	\centering
	  \includegraphics[width=130mm]{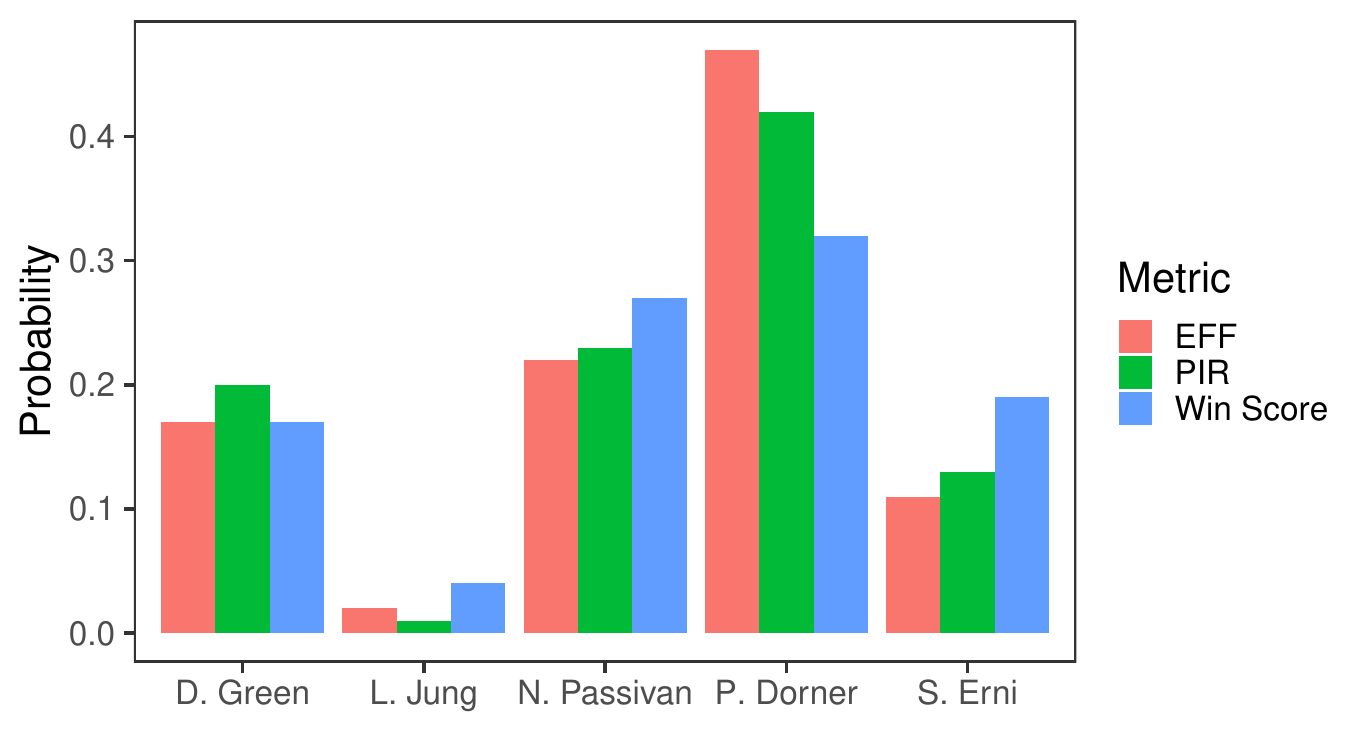} 
	 	\caption{Estimated probabilities of the inclusion of players Dejon Green, Lucas Jung, Natalie Passivan, Patrick Dorner and Svenja Erni in the line-up team given that players Annabel Breuer, Correy Rossi, Dirk Passivan and Walter Vlaanderen  have already been chosen,  based on EFF, PIR and Win Score metrics.} \label{fig:prob_5th_player}
\end{figure}

The amount of information we can extract from our analysis could be highly relevant. A notable example may be the following. In Figure \ref{fig:inclusion_probabilities} the four players most likely to be selected for the line-up are identified: Annabel Breuer, Correy Rossi, Dirk Passivan and Walter Vlaanderen. Now, the question arises: which player from the Doneck Dolphins Trier team would be the ideal complement to complete the composition? This question can be addressed in a simple way as follows:
\begin{equation*}
    P(I_i \mid I_1, I_2, I_4, I_9, \mathcal D) = \frac{P(I_i, I_1, I_2, I_4, I_9  \mid \mathcal D)}{P(I_1, I_2, I_4, I_9 \mid \mathcal D)}, \text{ with } i=\{3, 5, 6, 7, 8\},
\end{equation*}

\noindent where $P(I_i \mid I_1, I_2, I_4, I_9, \mathcal D)$ represents the probability of player $i$ being selected given that players 1, 2, 4, and 9 have already been chosen. Figure \ref{fig:prob_5th_player} shows a bar plot with the estimation of these probabilities considering EFF, PIR and Win Score metrics.

Finally, upon examining Figure \ref{fig:prob_5th_player}, it becomes evident that Patrick Dorner is the player who best completes the group. It is important to note that this team, consisting of $\{${Annabel Breuer}, {Correy Rossi}, {Dirk Passivan}, {Patrick }{Dorner}, {Walter Vlaanderen}$\}$, corresponds to line-up $L_1$. Additionally, the case of Natalie Passivan deserves special attention. On one hand,   Figure \ref{fig:inclusion_probabilities} indicates that the probability of including this player in the best composition is one of the lowest. However, when considering the information provided by Figure \ref{fig:prob_5th_player}, this player emerges as the second-best option to complete the most optimal line-up. This demonstrates the valuable information that can be derived from the results obtained through our methodology.

\subsection{Model checking} 
Finally, in order to interpret how accurate the model fits the data, we compute the cross validated probability integral transformation (PIT) defined for the player $i$ at match $j$ as
\begin{equation}
    PIT_{ij} = P(y^{(pre)}_{ij}\leq \mathcal{D}_{ij} \mid \mathcal{D}^{-ij}).
\end{equation}
\noindent Here, $\mathcal{D}_{ij}$ represents the observation for player $i$ during match $j$, and $\mathcal{D}^{-ij}$ encompasses all observations except that one. Directly computing these values can be computationally intensive because we would need to approximate as many posterior distributions as observations. However, this is not a requirement, as the application of self-normalised importance sampling enables us to approximate the $PIT's$ using samples drawn from the posterior distribution $\pi(\boldsymbol{\theta}, \boldsymbol{b} \mid \mathcal{D})$ computed with the complete data $\mathcal{D}$, as discussed in \citet{Gelfand2} and \citet{Ntzoufras}. Theoretically, if the model had produced the observation $\mathcal{D}_{ij}$, $PIT_{ij}$ should follow a uniform distribution $\text{U}(0, 1)$.

\begin{figure}[ht]
	\centering
	  \includegraphics[width=110mm]{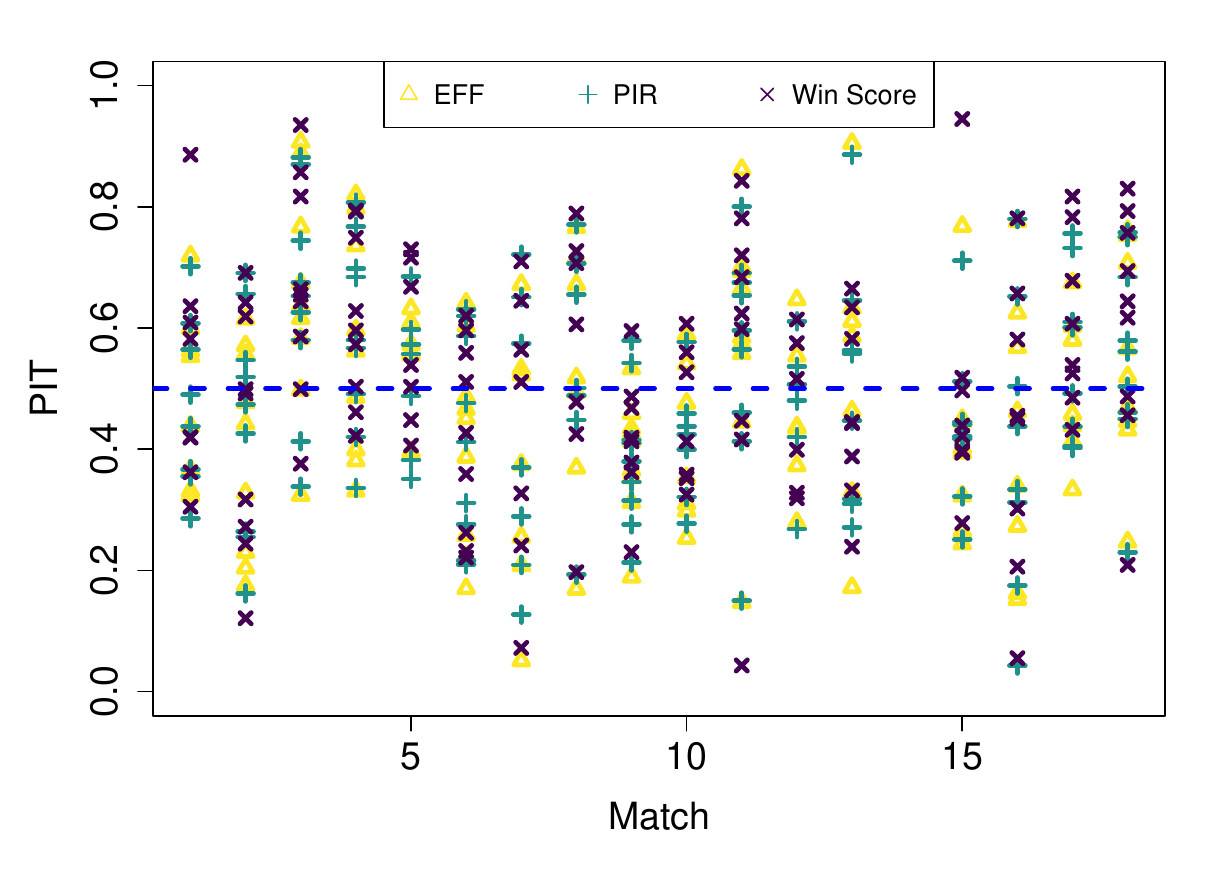} 
	 	\caption{Estimated cross validated $PIT$ values for the three metrics EFF, PIR and Win Score by match. The blue dashed line is the expected value $0.5$ of the theoretical distribution $\text{U}(0, 1)$.} \label{fig:PIT}
\end{figure}

Low $PIT$ values indicate model overpredictions, while high $PIT$ values indicate model underpredictions. Figure \ref{fig:PIT} illustrates the estimated $PIT$ values from the three models we computed. It can be observed that the majority of the $PIT$ values cluster around 0.5, indicating that this model fits the data reasonably well. {Nonetheless, they do not perfectly follow the uniform distribution $\text{U}(0, 1)$, likely because of two factors: the limited sample size of the data and  the fact that our model does not take the opponents into consideration, which obviously influence the players' performance}.

We can identify some unusual matches, where most of the associated $PIT$ values are either above or below 0.5. For instance, in the third match, there are clear instances of high $PIT$ values, while in matches 9 and 15, the majority of the values are below the blue line. What could this signify? It suggests that the model does not explain the players' performance well in those matches. However, it's important to note that the Doneck Dolphins won the third match with a final score of 81-57, corresponding to an unexpectedly large win margine. {In the same vein, matches 9 and 15 were lost with scores of 50-62 and 65-84, both at home, against BBC Munsterland and RSB Thuringia Bulls, respectively. These two teams finished above the Doneck Dolphins Trier in the final league standings.}

\section{Conclusions and future research}\label{sec:conclusion}

In this paper we have proposed a statistical methodology that allows to assess the best compositions for a wheelchair basketball team by using  previous data on the individual behaviour of the different players in the team.   Our proposal combines Bayesian longitudinal modelling, which analyses the performance data of players over time, with integer linear optimisation to determine the line-ups that maximise the performance of the team. This approach incorporates uncertainty throughout the entire process, resulting in a probability distribution for the optimal solution rather than a single solution to the optimisation problem.

The study we carried out using   data from a real wheelchair basketball team shows the mathematical and practical potential of our proposal. We clearly identify the most optimal composition of the Doneck Dolphins Trier team of the  German league, and illustrate how to calculate the possible compatibilities or incompatibilities between players on the court. While we may not have initially considered the potential absence of a player in a match due to a penalty or injury when calculating the optimal team line-ups, it is indeed possible to calculate the results under this particular circumstance. This can be achieved by obtaining the posterior probability of each alignment, denoted as $L_k$, while specifically taking into account those team line-ups that do not include the absent player.

In this research, we have exclusively focused on metrics that encompass positive actions and subtract negative actions, but it could be complemented with other types of metrics, such as those related to defense, as described by \cite{oliver2004basketball}. In addition, metrics based on possession-based statistics, such as offensive efficiency or defensive efficiency, as used in the NBA, can also be considered.

To conclude, this methodology has the potential to be immensely valuable for coaches and management teams, providing valuable insights into the ideal compositions for wheelchair basketball teams in upcoming matches. Furthermore, it can be readily applied to other team sports that cater to individuals with physical disabilities and incorporate  PCPS's within their regulations. This includes powerchair football, powerchair hockey, and wheelchair rugby among others. Moreover, this methodology can also be adapted for other team sports that do not have  PCPS's, by replacing the specific restrictions associated with the PCPS's rules with realistic constraints present in sports such as football, rugby, basketball, and others.

\section*{Acknowledgements}
Gabriel Calvo's research was partially funded by the ONCE Foundation, the Universia Foundation, and the Spanish Ministry of Education and Professional Training, grant FPU18/03101. Carmen Armero and Gabriel Calvo's research was partially funded by the Spanish Research project Bayes$\_$COCO (PID2019-106341GB-I00) from the Ministry of Science and Innovation Grant.

\bibliographystyle{apalike}
\bibliography{bibliography}

\end{document}